# The influence of dielectric properties on van der Waals/Casimir forces in solid-liquid systems

P.J. van Zwol, G. Palasantzas*, J. Th. M. De Hosson

Department of Applied Physics, Materials innovation institute M2i and Zernike Institute for Advanced Materials, University of Groningen, 9747 AG Groningen, The Netherlands

**Abstract**

In this article we present calculations of van der Waals/Casimir forces, described by Lifshitz theory, for the solid-liquid-solid system using measured dielectric functions of all involved materials for the wavelength range from millimeters down to subnanometers. It is shown that even if the dielectric function is known over all relevant frequency ranges, the scatter in the dielectric data, can lead to very large scatter in the calculated van der Waals/Casimir forces. Especially when the liquid dielectric function becomes comparable in magnitude to the dielectric function of one of the interacting solids, the associated variation in the force can be up to a factor of two for plate-plate separations *5-500 nm*. This corresponds to an uncertainty up to *100%* in the theory prediction for a specific system. As a result accuracy testing of the Lifshitz theory under these circumstances is rather questionable. Finally we discuss predictions of Lifshitz theory regarding multiple repulsive-attractive transitions with separation distance, as well as nontrivial scaling of the van der Waals/Casimir force with distance.

Pacs mumbers: 78.68.+m, 03.70.+k, 85.85.+j, 12.20.Fv

Corresponding authors: g.palasantzas@rug.nl ; petervanzwol@gmail.com

# I. Introduction

Forces between solid surfaces immersed in liquid have been intensively studied in colloid physics [1-19], where often water is used as the intervening medium [4,5]. Electrostatic double layer forces are present in many studies, and dispersion van der Waals (vdW) or Casimir forces are often approximated by the Hamaker theory [3], which states that dispersion forces scale with distance as $F \sim d^{-2}$, e.g., in the sphere-plane configuration, at short separations. If, however, higher accuracy is necessary for the calculation of dispersion forces, one has to employ the Lifshitz theory [2]. In that case, the knowledge of the dielectric function is necessary for frequencies that span several orders of magnitude, e.g., from the far-infrared to extreme ultraviolet regime. For many materials this requirement is problematic since only limited or even no dielectric data are available. As a result many researchers today still use the Hamaker theory [3], or the Ninham-Parsegian oscillator models [6] built from rather limited measured dielectric data [7-11].

Generally it is assumed that the Lifshitz theory accurately predicts the magnitude of dispersion forces in liquids when the correct dielectric data are incorporated. However, besides the many approximations or uncertainties in the available dielectric data, precise validation of the Lifshitz theory in liquids (i.e. without fitting theory to force measurements) only recently started [17-19]. Therefore, knowledge of the intrinsic uncertainty of the force calculations is necessary. Another source of uncertainty is that dielectric data are obtained by different groups using different techniques [12-14]. In addition, variation in the dielectric data can also been attributed to the quality of the material preparation. At any rate, variation in dielectric data results into an uncertainty in the calculated forces using the Lifshitz theory, unless the dielectric properties of the specific materials used for the force measurements are also accurately measured over the full frequency range [12].

Finally, it is known that the Lifshitz theory can predict relatively weak or repulsive dispersion forces between surfaces immersed in liquids [15-19]. Repulsive forces arise when for the dielectric functions $\varepsilon(i\zeta)$ (computed by the Kramers-Kronig relation at imaginary frequencies) of the surfaces 1 and 2 we have $\varepsilon_1(i\zeta) > \varepsilon_{liquid}(i\zeta) > \varepsilon_2(i\zeta)$ for all frequencies $\zeta$. The force is attractive when this is not the case for all frequencies. When $\varepsilon_{liquid}(i\zeta) = \varepsilon_{surface}(i\zeta)$ for all frequencies $\zeta$, then there is no force. Although in reality this will never be the case, for some finite frequency regime the dielectric function of the liquid is larger than that of one of the surfaces, $\varepsilon_{liquid}(i\zeta) > \varepsilon_{(1\ or\ 2)}(i\zeta)$, whereas in another regime the inverse may occur. This situation is exactly the case for the system silica-liquid-gold for which we will present Lifshitz theory calculations for multiple liquids with various degrees of knowledge of the dielectric functions.

## II. Lifshitz theory for force calculations

For the force calculations we will use Lifshitz theory without the temperature correction [2]. At measurable plate-plate separations (<*500 nm*) the finite temperature corrections are insignificant if accuracy above the 1 % level is sought [20]. Because the dielectric response of materials can change with temperature specifically in the microwave regime [21], the dielectric data for all materials presented here are obtained at room temperature where also the force measurements are performed [17-19]. For sake of clarity we will also ignore roughness corrections [22]. Furthermore, the vdW/Casimir energy between real parallel flat mirrors with area A, separated a distance L, with reflection coefficient $r_{ij}(q,\zeta)$ is given by

$$E_{pp} = \frac{\hbar}{2\pi} A \sum_P \int \frac{d^2q}{(2\pi)^2} \int_0^\infty d\zeta \ln[1 - r_{31}^p r_{32}^p e^{-2\kappa_3 L}] \qquad (1)$$

Where $q$ is the transverse wave vector (and q=|**q**|) and $p$ distinguishes the two polarizations of the electromagnetic field (*TM* and *TE*). The reflection amplitudes $r^{p,s}_{3i}$ are given by the Fresnel reflection coefficients

$$r^{TE}_{3,i} = \frac{\kappa_i - \kappa_3}{\kappa_i + \kappa_3}, \quad r^{TM}_{3,i} = \frac{\kappa_i \varepsilon_3(i\zeta) - \kappa_3 \varepsilon_i(i\zeta)}{\kappa_i \varepsilon_3(i\zeta) + \kappa_3 \varepsilon_i(i\zeta)}, \quad \text{and} \quad \kappa_i = \sqrt{q^2 + \frac{\varepsilon_i(i\zeta)\zeta^2}{c^2}} \quad (2)$$

where $\varepsilon_{1,2}$ are the dielectric functions for the surfaces, and $\varepsilon_3$ is the dielectric function for the liquid. We obtain the force in the plate sphere setup using the Derjaguin approximation for the force $F=2\pi R_{sph}E_{pp}$ with $E_{PP}$ defined above and $R_{sph}$ the sphere radius. The sphere radius used in all calculations is $R_{sph}$ = 9 µm since we have used these spheres for our force measurements.

The reflection amplitudes depend on the dielectric function at imaginary frequencies $\varepsilon(i\zeta)$, which is obtained by the Kramers-Kronig analysis of the measured imaginary part $\varepsilon''(\omega)$ of the real dielectric function $\varepsilon(\omega)$

$$\varepsilon(i\zeta) = 1 + \frac{2}{\pi} \int_0^{+\infty} d\omega \frac{\omega \varepsilon''(\omega)}{\omega^2 + \zeta^2}. \quad (3)$$

For all our materials, we obtained the dielectric function at imaginary frequency by direct numerical integration of the imaginary part of the dielectric function. This procedure is allowed because the dielectric data are available over a sufficiently large range of frequencies.

Since we will frequently refer to the Ninham-Parsegian oscillator models [6] for comparison to real dielectric data, their definition will be briefly presented here. In fact, oscillator models are constructed to represent the dielectric function at imaginary frequencies, in the case when there is little or no knowledge of the dielectric function of a substance. The form of the oscillator model is given by

$$\varepsilon(i\zeta) = 1 + \sum_i \frac{C_i}{1 + (\zeta/\omega_i)^2}. \quad (4)$$

The coefficient $C_i$ is the oscillator strength at a given (resonance) frequency $\omega_i$. Typically two or three oscillator functions are used, namely, an ultraviolet (UV) part, infrared (IR) part, and sometimes a microwave (MW) part (though its form can be slightly different from the typical dependence $\sim(\zeta/\omega_i)^2$ in Eq. (4) [8]). The sum of $C_i$'s should always satisfy the relation $\sum C_i = \varepsilon_0 - 1$ with $\varepsilon_0$ the static dielectric constant of the substance.

### III. Accuracy of force calculations for specific systems

It is known that the variation of the dielectric functions, e.g., for gold, can be rather large because of defects and grains (depending on the manufacturing process) in the material. This may lead to a variation in vdW/Casimir force of *5-15%* as calculations in terms of the Lifshitz theory for gold surfaces in air indicated [12]. Obviously for other materials we may expect similar behavior. In fact, for silica the variation in the dielectric functions are due to variations in density and water content [13]. For liquids, besides system temperature, also the purity plays a significant role on their dielectric response [14]. Therefore, assuming that the Lifshitz theory, Eq. (1) is accurate, which is not uncontroversial since there may be issues as to the range of validity of the underlying electromagnetic stress tensor inside material media that goes into the derivation, scatter in the dielectric data, which is the input for the force calculations, is the only significant source of error.

For gold, silica, and water the dielectric functions are well known and measured by various groups. We will use two sets of data for gold as published in ref [12]. Also two sets of data will be used for silica as obtained by different groups [13]. Finally, for water we will use the data of Segelstein [14], and an *11-order* oscillator model [11] that has been fit to different sets of data [7,9]. All the dielectric data are shown in Fig. 1, which depicts the significant scatter in dielectric data for all materials.

Fig. 2 shows the calculated forces using Lifshitz theory for the different sets of dielectric data. Here we would like to emphasize the following. Although in the case of silica one of the data set has significantly less data points in the visible and near infrared (IR) range, this is not the reason for the different outcome in the force calculations. This is because the major adsorption peaks (UV and IR range), which define the dielectric function at imaginary frequencies and the calculated forces, contain sufficiently enough data points. Furthermore, the scaling of the force with distance in Fig. 2c is defined as *scaling=|(-log$_{10}$(|F$_i$|)+log(|F$_{i-1}$|)) / (log(X$_i$)-log(X$_{i-1}$))| wh*ere $F_i$ and $X_i$ represent force and separation for successive data points *i*, respectively.

Surprisingly for the gold-water-silica system the scatter in the dielectric data leads to scatter in the calculated forces, which reaches the level of *60%* for separations *L<500 nm*. This scatter is much larger than the variation in the force between gold surfaces in air. Therefore, unless the dielectric properties of the specific samples used in force measurement are measured over a very large frequency range, any comparison of force measurements in liquid to Lifshitz theory predictions is only meaningful as a rough estimate (i.e. within ~*50%* error bars).

The reason for these dramatic effects originates from the fact that the magnitude of the dielectric functions at imaginary frequencies for water and silica are similar for most frequencies. Indeed, in the range *$10^{-2}$-$10^2$ eV* the difference is of the order of *30%* or less, which is in fact similar in magnitude to the scatter of the measured dielectric functions. To illustrate this effect we present calculations for silica-water-silica and gold-water-gold in Fig. 3. Besides the fact that in the case of silica the force has decreased to below force resolution level of a typical AFM colloid probe system (the force is less than *10 pN* at separations *L≥10 nm*), the scatter in the force calculations becomes very large, e.g., *100%* for silica compared to *18%* for gold. Therefore, the accuracy of the force calculations for the systems described in [17] and [18] is rather low, while in [19] the accuracy of the force calculations can be better. Nonetheless, in all these cases [17-19] oscillator models were used for the liquids, which contribute to inaccuracy of the calculations as will be shown in the following.

### IV. Oscillator models for representing the dielectric function

#### *(a) The case of water*

The accuracy of the force calculation will further deteriorate if poorly constructed oscillator models are used to represent the dielectric response of liquids. The problem for most liquids, except water, is that the dielectric function is not well known [6-11]. As a consequence it is understandable that oscillator functions [6, 7] are constructed from the limited available measured dielectric data to obtain an approximate dielectric function at imaginary frequencies for a specific liquid [8]. For example, sometimes the dielectric strength measured at only a few fixed frequencies (e.g., the sodium D line) were used to determine the strength of an oscillator [8].

Since for water the dielectric function is known, we can make a comparison to oscillator functions provided in [10]. Fig. 4 shows that the difference with the real data is very large. In particular the dielectric strength in the UV regime is underestimated, and in the IR regime it is rather overestimated. The result for the force in the non-retarded (or vdW) regime is not large. Indeed, the force is about *25%* times larger for separations *< 10 nm*. However, in the retarded regime the force becomes even repulsive for separations *L>25 nm*. It should be noted, however, that the oscillator models in [7, 9] reproduces the dielectric data by Segelstein [14] rather well.

One can argue that the other oscillator model as constructed by van Oss et al. [10] (where the same parameter values for water can be found in [6, 8]) is highly inaccurate. Nonetheless, probably the large difference in the dielectric strength in the IR regimes can be attributed to average values obtained from a limited data set around the IR regime. Notably, in those days there was not much focus on IR parameters since the UV parameters were of paramount importance for the non-retarded Hamaker constants. This is also one of the reasons why the oscillator model in [6, 10] still works relatively well below *10 nm*.

### *(b) The case of ethanol*

The measured dielectric function for ethanol presented here was constructed over a very wide range of frequencies from real measured data provided in existing literature (Fig. 5). We will compare force calculations using this data to force calculations based on the oscillator models for ethanol that have been used for comparison to measurements in [19] where the model of Milling et al. [16] was used for force calculations. For the measurements in [18] the three oscillator model of van Oss et al. [10] was used for the force calculations.

The dielectric function for ethanol was constructed as follows. The millimeter range was taken from [23], the micrometer range from [24], the mid infrared range from [25], the near infrared range from [26], and the ultraviolet and x-ray range from [27]. Only in the visible and in the far infrared range some data are still lacking. However in the visible range ethanol is transparent (very low extinction coefficient, which can be assumed to be zero in our case), and therefore this part will not contribute to the dielectric function at imaginary frequencies. The far-IR regime may have some features, but we do not expect order of magnitude changes from the linear interpolation in this range. In any case the lack of knowledge of the dielectric function in this range will not give a large contribution for the calculated forces in the ranges considered here (*<500 nm*). On the other hand, the UV part will have the predominant role in the calculation of the force at smaller ranges, and for this part we have reliable data [27]. Although in the literature a significant spread exists in the measurements of dielectric data $\varepsilon(\omega)$ for ethanol [27], we derived a fairly accurate dielectric function at imaginary frequencies $\varepsilon(i\zeta)$ by means of Eq. (3).

The results of the calculated forces for silica-ethanol-gold are shown in Fig. 6. Although the oscillator model of Milling et al. [16] for ethanol predicts repulsive forces over all ranges, attractive forces were measured [18]. The three oscillator model of van Oss et al. [10] shows a better agreement with the forces obtained from measured dielectric data for ethanol, but it predicts slightly weaker forces. The force measurements in [18] are in reasonable agreement with the calculated forces obtained with the full measured dielectric spectrum of ethanol (within *50 %* over most ranges, see also Fig. 10). In particular for this system the scatter in theory is of the order of *50-100%* or a factor of 2 when using different sets of measured dielectric data. Therefore, we can conclude that they are consistent at that level of accuracy.

In Fig. 7 data are shown for silica-ethanol-silica and gold-ethanol-gold. In the case of silica-ethanol-silica the oscillator model of Milling et al. [16] predicts forces which are more than an order of magnitude too large. For gold-ethanol-gold force measurements were found to be in agreement to the model of Milling et al. [16] in the range *20-100 nm* [19]. Therefore, it is worthwhile to compare this model to the real dielectric data for this system. It appears that when using this model in the range *20-100 nm* the forces are underestimated by *30-60 %* in the range *20-100 nm*. But given the large scatter in the measured forces in [19], we feel that this experiment was not accurate enough to be sensitive to these differences [19]. Moreover, also in that case we can conclude that the measurements in [19] are in reasonable agreement with the correct theory within roughly *30-50 %* uncertainty.

The poor quality of the oscillator model by Milling et al. [16] stems from the fact that they assumed that there was no significant absorption in the microwave regime. As a result they gave an IR oscillator with high strength because their model must satisfy $\sum C_i = \varepsilon_0 - 1$. This argument is unphysical since polar liquids such as water and ethanol absorb strongly in the microwave region described by Debye relaxation. The Debye term is included in the model of van Oss et al. [10], which gives the reason why their model gives somewhat better results. Therefore, it is of no surprise why Milling et al. [16] appear to have found erratic behavior (attractive forces) when they measured forces between teflon and gold surfaces in ethanol.

*(c) The case of bromobenzene*

Bromobenzene has recently been used for the measurement of repulsive forces between gold and silica surfaces [17]. Extremely weak repulsion was measured, indicating that the dielectric functions of bromobenzene and silica are very similar in magnitude. Once again an oscillator model of Milling et al. [16] was used for the calculations in [17]. Therefore, we will compare this model to measured dielectric data and other oscillator models. We have found measured dielectric data of bromobenzene in the MW [28] and IR [29, 30], and UV [31] regimes (Fig. 8). However we did not find sufficient data in the UV regime for the complete dielectric function. The discussion will therefore be concentrated on the MW and IR contribution of the dielectric function on the calculated forces.

Another oscillator model also exists for bromobenzene [32]. The two different oscillator models of Milling et al. [16] and Drummond & Chan [32] differ only significantly in the IR regime (Fig. 9a, inset). Once again Milling et al. ignored the MW absorption leading to a stronger IR absorption (or oscillator strength) [16], which is the reason for the difference between the two models. From Fig. 8, however, it is clear that there is significant absorption in the MW regime which cannot be ignored. The measured absorption in the IR regime on the contrary is rather low (on average much lower than that of water and ethanol). One can use the Kramers-Kronig relation to obtain the contribution of the IR part to the dielectric function at imaginary frequencies. It appears that the model of Drummond et al. [32] gives very similar results to the measured data in the IR regime. This is not a surprise since they have used the measured absorption in the microwave regime, as well as to obtain an estimate of the IR absorption [32]. The strength of the IR oscillator is therefore weak, and therefore the model of Milling et al. [16] is unreliable.

However, a problem still remains. The model of Drummond et al. [32] predicts an extremely weak repulsion, which is about twenty times weaker (as Fig. 9 shows) than the measured forces in [17]. In addition, the oscillator strength in the IR and MW regimes of the model of Drummond et al. [32] are known to be correct, and therefore the strength of the oscillator in the UV regime must also be correct by definition since the sum of all strengths must satisfy the relation $\sum C_i = \varepsilon_0 - 1$ [8]. However the shape of the real dielectric function at imaginary frequencies depends on the local structure of the measured real dielectric behavior, and this is not known in the far UV regime. For example the frequencies used in oscillator models are chosen to be at some resonance for a specific electronic bond [8]. Since there are many bonds in a molecule such as bromobenzene, there are multiple resonance frequencies to choose from.

This leads us to conclude that the theory prediction for the silica-bromobenzene-gold system is, at present, uncertain, and further studies are necessary despite the agreement between the force measurements in [17] and the theory prediction using the oscillator model of Milling et al.[16].

## VI. Theory and experiment

For the sake of completeness we now compare all calculations to force measurements for the gold-ethanol-silica system [18]. It is important to know that these measurements were done expecting repulsive forces based on the model of Milling et al. [16]. Multiple dielectric data were used for gold and silica. For ethanol the dielectric data are varied by *15%* in the UV regime (see [27] for the variation in absorption measurements in the UV regime). For all materials there is roughly *10-30 %* variation in the measured dielectric data. When oscillator models are used for ethanol, then we use only one data set for gold and silica.

From figure 10 it is clear that the measured force data follow the full theory showing similar scaling with distance. However, the scatter in the theory is very large, and only accurate within a factor of two (or equivalently *100 %*). When the model of van Oss. et al.[10] was used for ethanol, the discrepancy between theory and measurements becomes larger at the smallest and largest separations. Note that any discrepancies between the different calculations and the force measurements at the smallest ranges (*<10nm*) can be explained by the uncertainty in distance of the measurement (~1nm). At the largest ranges the error in distance leads to smaller uncertainty in the measured forces. Clearly the oscillator model of Milling et al. [16] gives a completely wrong prediction of the force.

Summarizing the results for water, ethanol and bromobenzene, we can infer that oscillator models are not suitable for precision calculations using the Lifshitz theory. They can even lead to completely wrong predictions. The prime reason for their use was the ability to construct dielectric functions at imaginary frequencies, while only having limited experimental data. In that sense, these models are only useful to obtain a rough estimate of the force, e.g. within an order of magnitude, unless substantial measured dielectric data are used to determine the specific strength of each oscillator.

**VI. Repulsive-Attractive transitions with distance**

Finally, we will discuss the possibility of attractive-repulsive force transitions. It is known that when the dielectric function of the liquid becomes comparable to the dielectric function of one of the interacting solids, then very weak forces are predicted. However, with a high uncertainty in the magnitude of the force due to uncertainty in the measured dielectric data. The same is true for the actual shape of the force since the dielectric functions can overlap at some places and cross each other. Highly complex behavior of the force is then observed. We will illustrate this using oscillator models for p-xylene from [16], and bromobenzene by (see Fig. 11a inset) [32].

It is clear from the discussion above that these calculations should be regarded as only indicative, since the knowledge of the dielectric functions of these liquids is incomplete. Obviously in the case of bromobenzene measurements indicate short range repulsion [17]. Thus it is likely that the UV part of the oscillator function used here for bromobenzene [32] is not correct, while the IR part is correct as it was discussed previously. The results are shown in Fig. 11. While the forces are rather weak, the force switches sign twice with distance in the case for bromobenzene. However, in the case of p-xylene, which has a rather similar dielectric function compared to that of bromobenzene, but only slightly smaller, the force does not switch sign. Instead it shows a highly non trivial scaling of the force with surface separation. First the scaling increases from *2* in the non retarded regime to *4* at *40 nm*, decreases to roughly *2* again at *150 nm*, and then it increases again to about *3* in the retarded regime. Which is rather different from normal 'retardation' effects were the scaling of the force goes slowly from *2* (non retarded at *L<10 nm*) to 3 (retarded regime *L>10nm*) as in Fig. 2c.

Thus, as becomes apparent from these two examples, very slight changes in the dielectric function can give a completely different force-distance behavior. Note that indirect evidence of the sign switch can be seen in an experiment from a strong deviation of scaling on a log-log scale near the transition point (Fig. 11c, d). The ideal case for the detection of this attraction-repulsion transition with distance would be a solid that has strong UV adsorption and weak IR absorption, and a liquid with weak UV absorption and strong IR absorption.

## VII. Conclusion

We have shown that the scatter in the measured dielectric data of solid materials and liquids can lead to large scatter of up to a factor of two and possibly more in the calculated vdW/Casimir force as obtained by Lifshitz theory. This particularly happens in the case when the dielectric functions of the liquid and one of the interacting surfaces become comparable in magnitude. As a result in this case the agreement between force measurements and Lifshitz theory cannot be better than ~*50%*, unless the dielectric functions of the materials involved in the force measurements are also measured.

Moreover, if oscillator models are used to represent the liquid dielectric function, the uncertainty in the calculated forces can increase up to an order of magnitude. This is because the oscillator models were constructed from very limited dielectric data, and as a result can not be used for accurate testing of Lifshitz theory. Obviously measurement of the full dielectric functions of the liquids and surfaces is required as input in Lifshitz theory calculations, before any claims of accuracy can be made. The present state of the field is that Lifshitz theory, for solid liquid systems, has only been verified to be in agreement with experiments within roughly *50 %* accuracy in the best case. Nonetheless, Lifshitz theory can predict interesting effects such as multiple sign switches with distance and non-trivial scaling of vdW-Casimir forces with distance, in regimes accessible by direct force measurements using common colloid probe atomic force microscopy.

## Acknowledgments

The research was carried out under project number MC3.05242 in the framework of the Strategic Research programme of the Materials innovation institute M2i, (the former Netherlands Institute for Metals Research or NIMR). Financial support from the M2i is gratefully acknowledged. We would like to acknowledge useful discussions with V. A. Parsegian, J. Munday, F. Capasso, and V. B. Svetovoy.

**Figure Captions**

**Figure 1.** (Color online) **(a)** Raw dielectric data of the materials obtained from different references. **(b)** Dielectric functions at imaginary frequencies. **(c)** is a magnification of **(b)** showing better the differences between water and silica. The solid and dashed lines for silica and gold are 2 different sets of optical data. For water the solid line are the data from Ref. [14], and the dashed line is an 11-order oscillator model which has been fitted to a different set of optical data.

**Figure 2.** (Color online) Calculations of the Casimir force for the different sets of data. The solid line is the reference line. Circles, squares, triangles and stars are calculations when we interchange one set of dielectric data with another one for one or more materials. (The upper two graphs show the force in linear **(a)** and log scale **(b)** with distance. **(c)** Absolute value of the scaling of the force with distance. Clearly squared to cubed scaling, e.g. from non retarded to retarded regimes is observed. **(d)** The relative difference in force as compared to the reference force in percent.

**Figure 3.** (Color online) **(a)** Lifshitz theory calculations for gold-water-gold and silica-water-silica. **(b)** Relative difference in percent of the force when different measured dielectric data (as shown in Fig 1) for gold, and glass is used. For gold the effect on the force is small, while for silica it is very large because its dielectric function is similar to water.

**Figure 4.** (Color online) Lifshitz theory for gold-water-silica. Here we calculate the force using real measured data by Segelstein, and oscillator models, one from ref. 10 (van Oss et al), and one from refs 7, 9 and 11. The force is shown in linear scale **(a)**. The inset in (a) shows the dielectric functions of all materials and the various oscillator models for water. In **(b)** we show relative differences in percent (%) for the various oscillator models as compared to the data in [14].

**Figure 5.** (Color online) The extinction coefficient for ethanol over a wide range of frequencies from refs. 23-27. With the Kramers-Kronig relations the dielectric function at imaginary frequencies can be obtained from this graph. For comparison we also show the extinction coefficient for water [14]. All data are shown on log-log and semi log scale.

**Figure 6.** (Color online) **(a)** Calculated forces for gold-ethanol-silica using measured dielectric data for gold, silica and ethanol. We compare the latter to force calculations based on the oscillator models of the dielectric function for ethanol from [10] and [16]. The inset shows the dielectric function of gold silica and ethanol together with those of the two oscillator models for ethanol. **(b)** The relative difference in force for the two oscillator models as compared to the forces obtained from measured dielectric data (used as reference: $F_o$). The model of van Oss et al. [10] is relatively accurate (within a factor 2 for separations < 100 nm), while the model of Milling et al. [16] predicts repulsive forces.

**Figure 7.** (Color online) **(a)** Calculated forces for gold-ethanol-gold (yellow lines) and silica-ethanol-silica (black lines). The solid data are obtained with measured dielectric data for ethanol. The dashed line is obtained with the oscillator model of Milling et al. [16]. The differences are extremely large in the case of silica were the oscillator model predicts forces

more than an order of magnitude too large. **(b)** For gold-ethanol-gold the situation is better, but still the forces are underestimated by a factor 10-80 %.

**Figure 8:** (Color online) Extinction coefficients for bromobenzene taken from refs. 28-31, compared to water on log-log and semi-log scales. From this data we can say that the oscillator function as obtained in [32] is correct in the IR regime.

**Figure 9.** (Color online) **(a)** Calculated forces for gold-bromobenzene-silica using measured dielectric data for gold, silica and oscillator models for bromobenzene obtained from [16, 32]. The inset shows the dielectric function of gold and silica together with the two oscillator models for bromobenzene. **(b)** The relative difference in force for the two oscillator models. The model of Drummond and Chan [32] is used as a reference because it is better in the IR regime. We stress that both models have almost the same UV oscillator strengths and frequencies, but these may not reflect the real dielectric response of bromobenzene in the UV regime. The force switches sign twice (attractive-repulsive-attractive) when using the model of Drummond et al. (see also Fig. 11)

**Figure 10.** (Color online) Force measurements (black squares and circles), plotted on log-log and semi-log scale for the gold-ethanol-silica system averaged over multiple spheres from [18], compared to theory using multiple dielectric data sets for gold, ethanol, and silica (Grey solid lines). A comparison with two oscillator models for ethanol is made, the one of van Oss et al. (red squares on a solid line) [10] and the one of Milling et al. (red circles on a solid line ) [16]. Clearly the measurements coincide at best with a set of the full theory, showing similar scaling with distance. Especially at large separations where the uncertainties in the measurement becomes smaller (because the error in distance is ~1 nm affecting strongly short seprations < *10 nm*) the agreement with the full theory is good, and becomes worse when using the respective oscillator models to represent ethanol.

**Figure 11.** (Color online) Force calculations for bromobenzene and p-xylene on semilog and log scales **(a-c)**. The liquids are represented by oscillator models (inset in **(a)**). Highly complex force curves are obtained when the dielectric functions start to overlap or cross each other. Multiple sign switches and nontrivial scaling with distance can result from this. While the forces are extremely weak, around a transition point a sign switch can be observed experimentally since there is a strong deviation from the normal scaling of the force with distance **(d)** this is quite visible on log-log scales **(b,c)**.

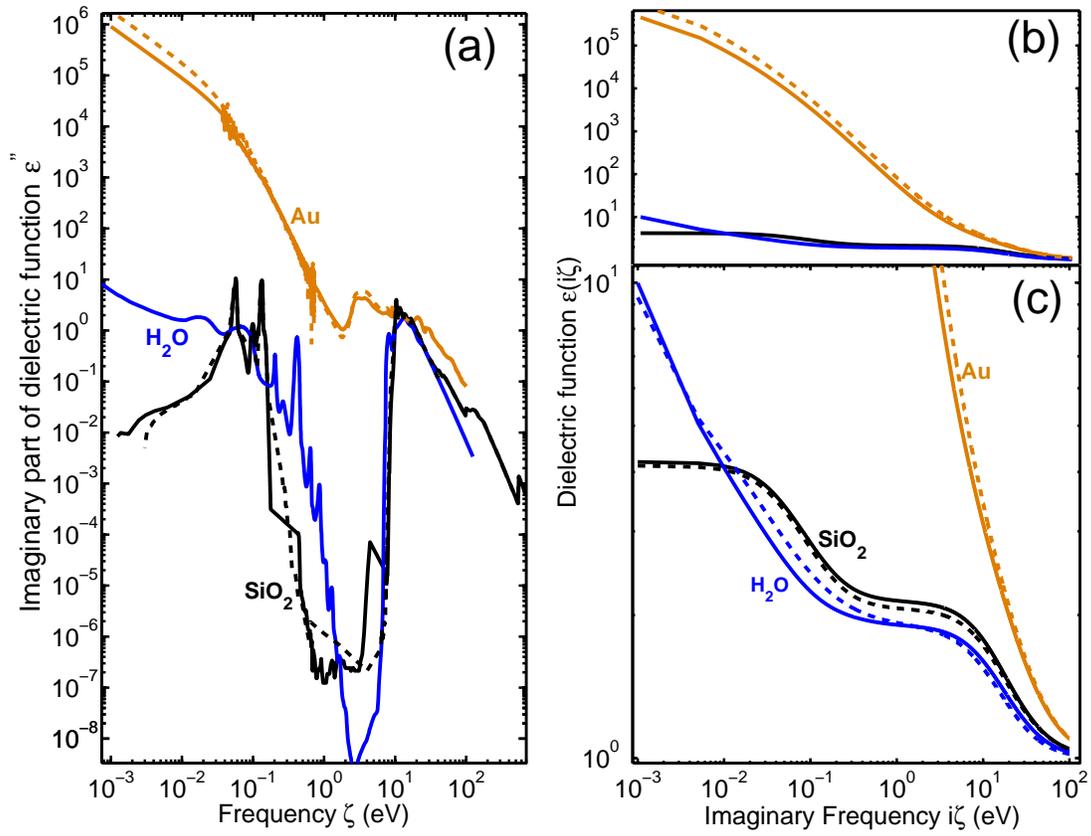

**Figure 1**

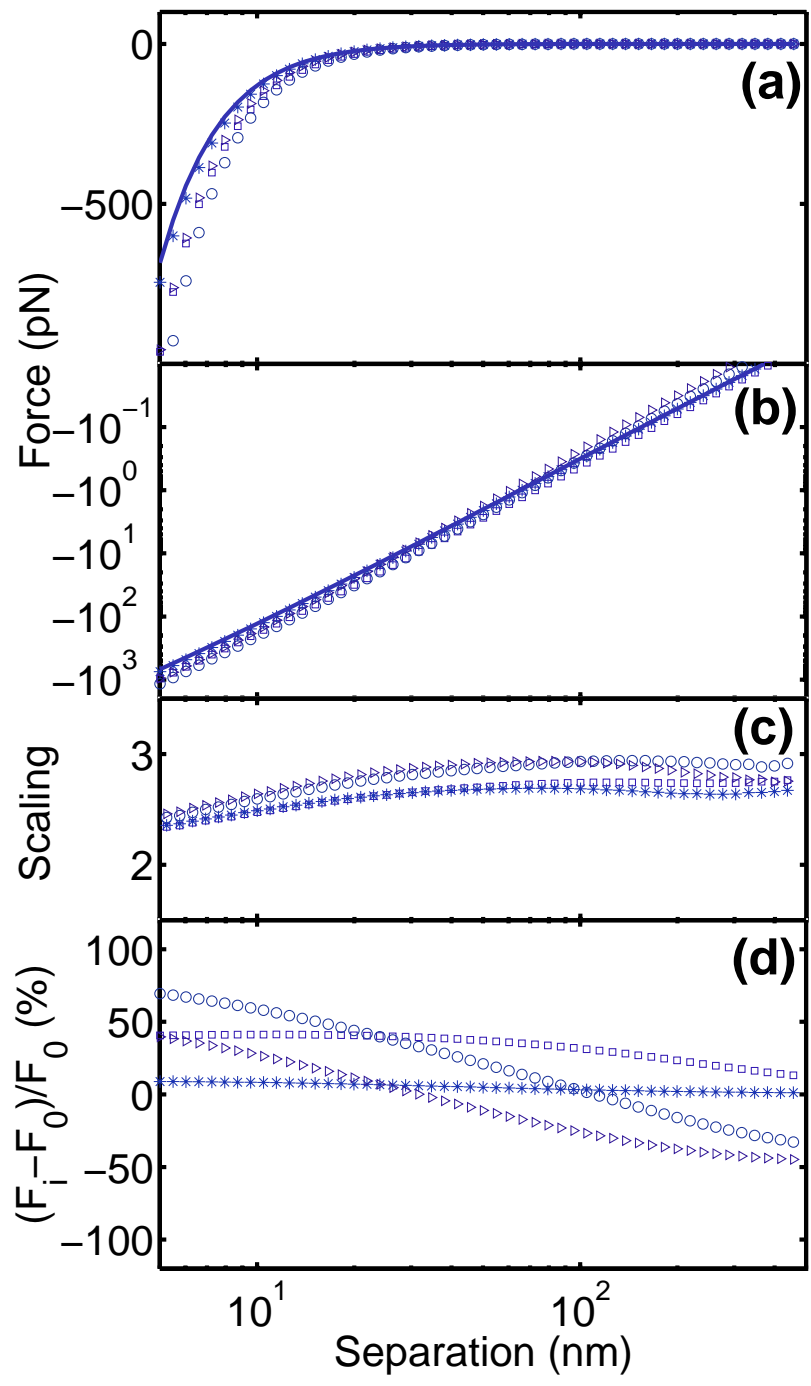

**Figure 2**

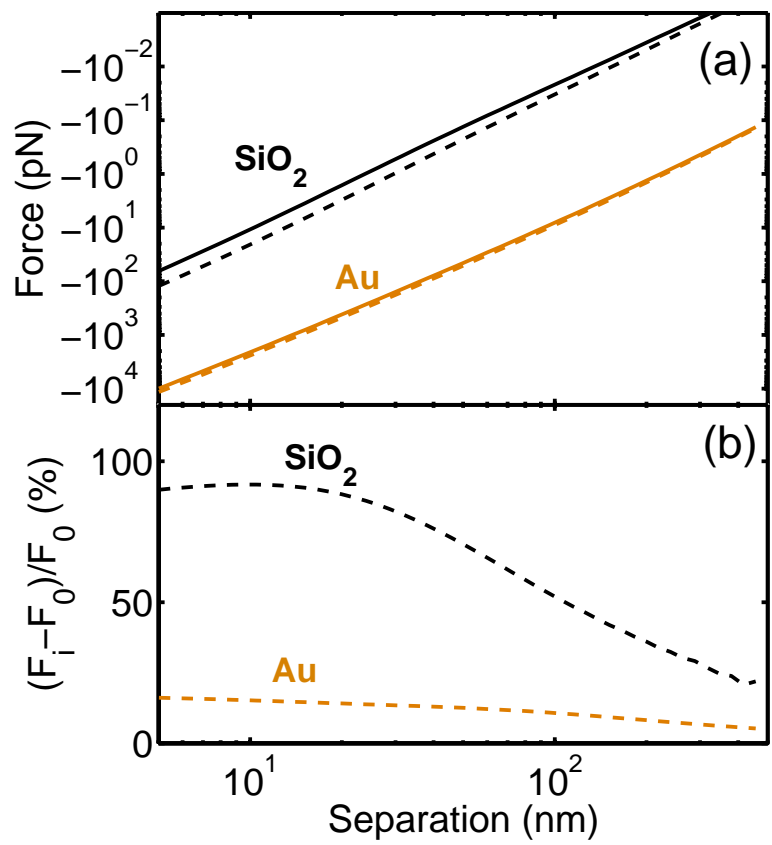

**Figure 3**

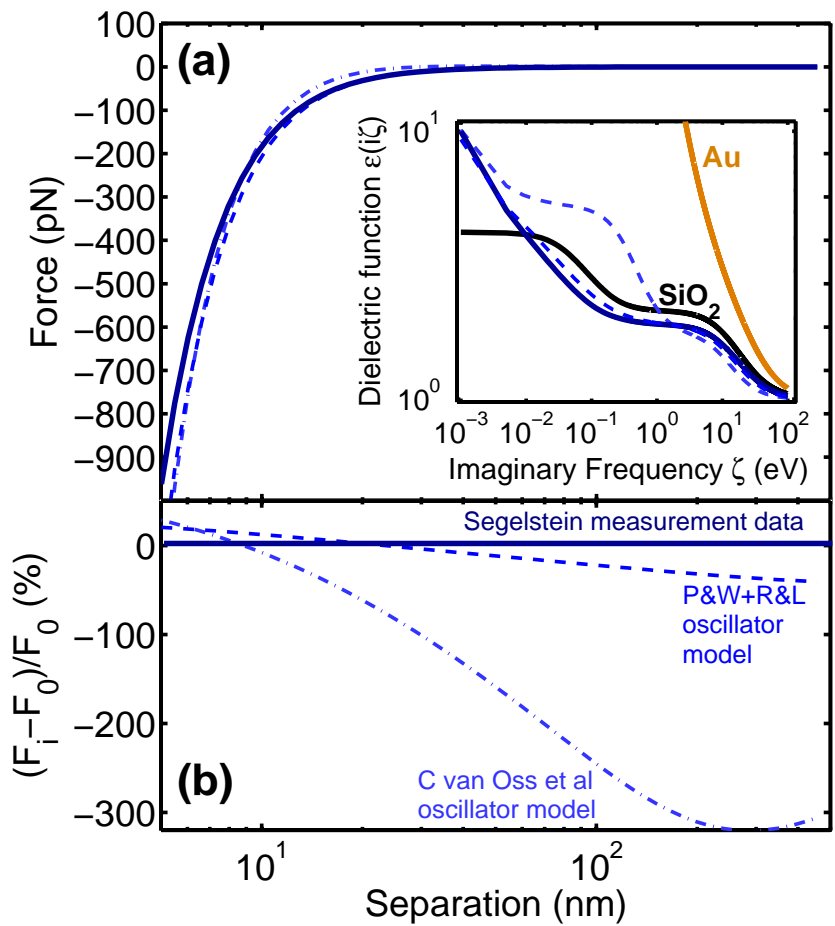

**Figure 4**

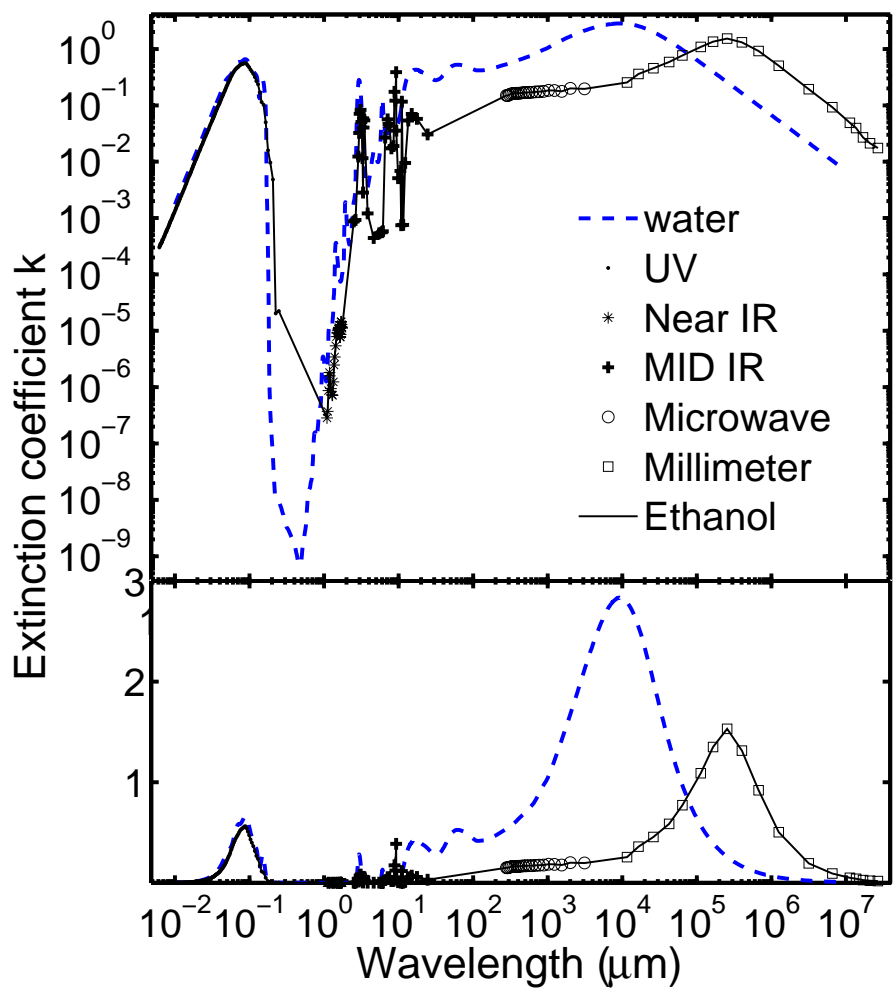

**Figure 5**

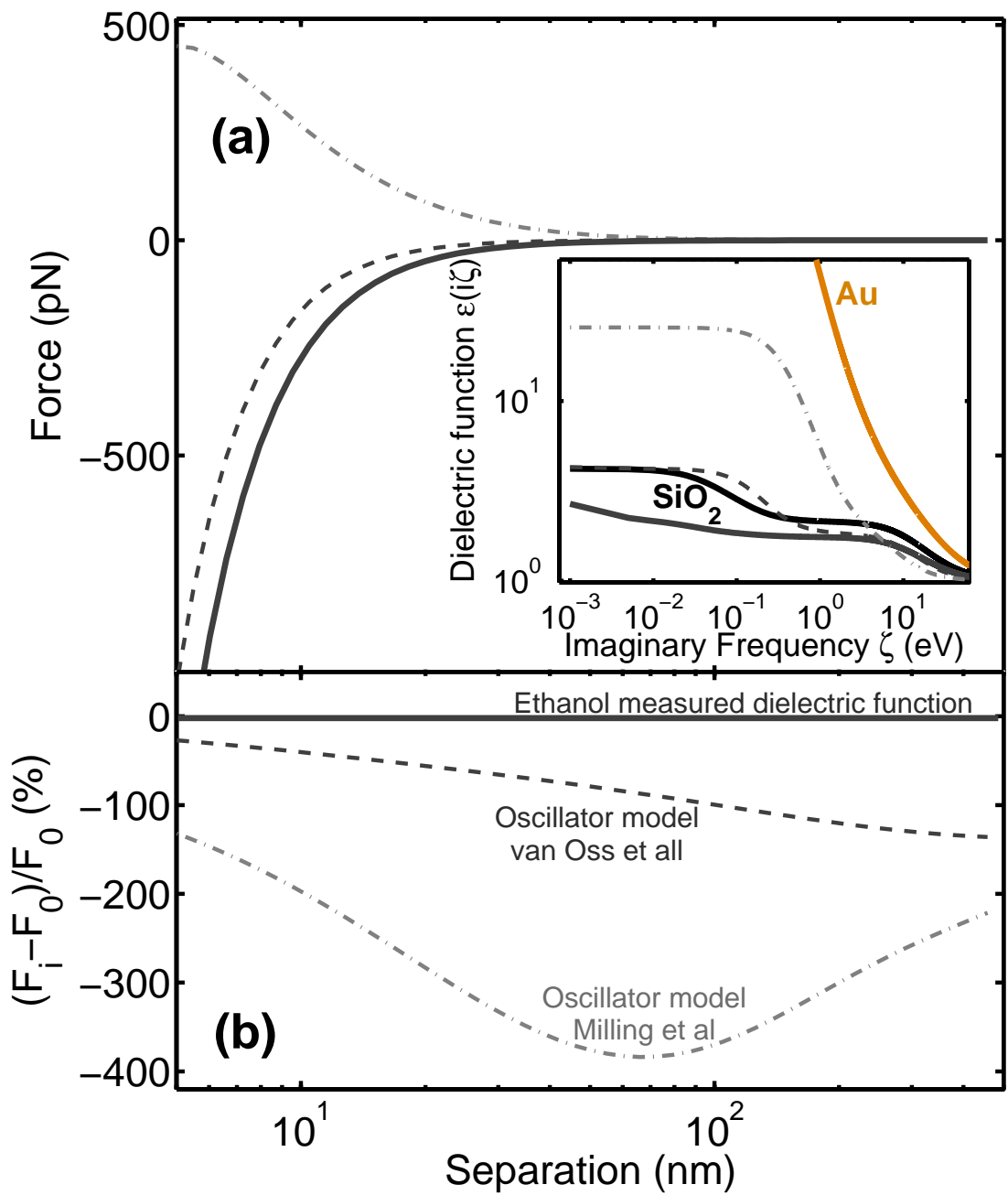

**Figure 6**

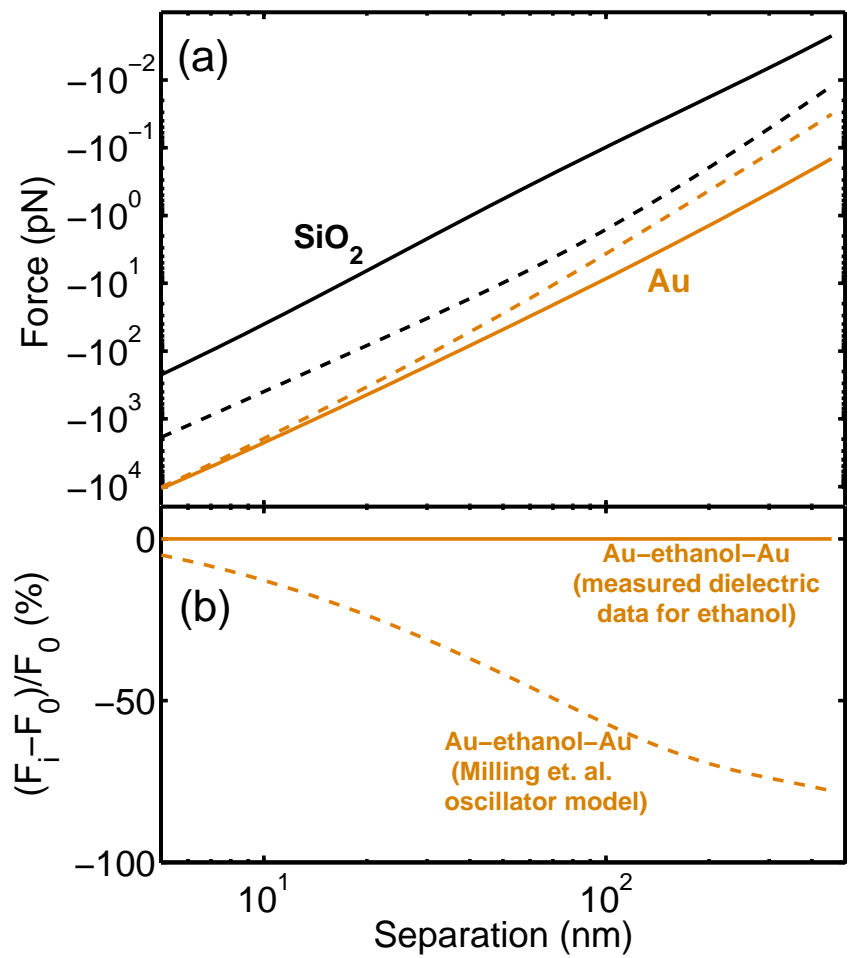

**Figure 7**

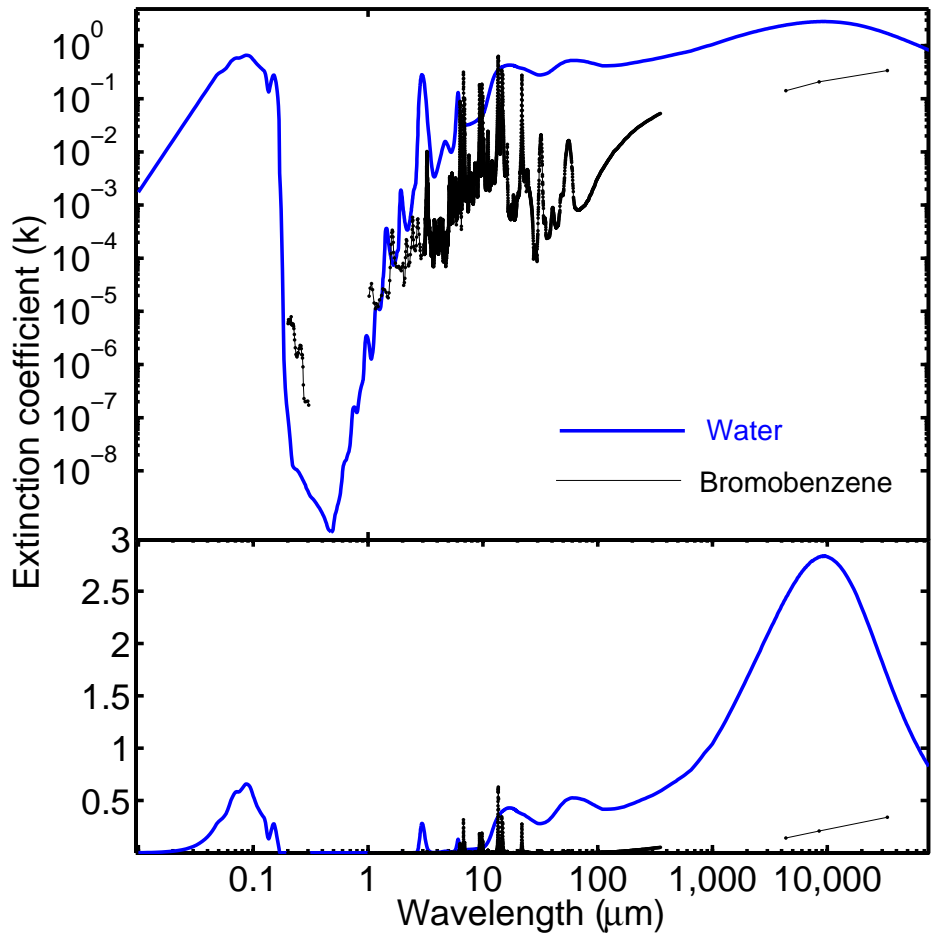

**Figure 8**

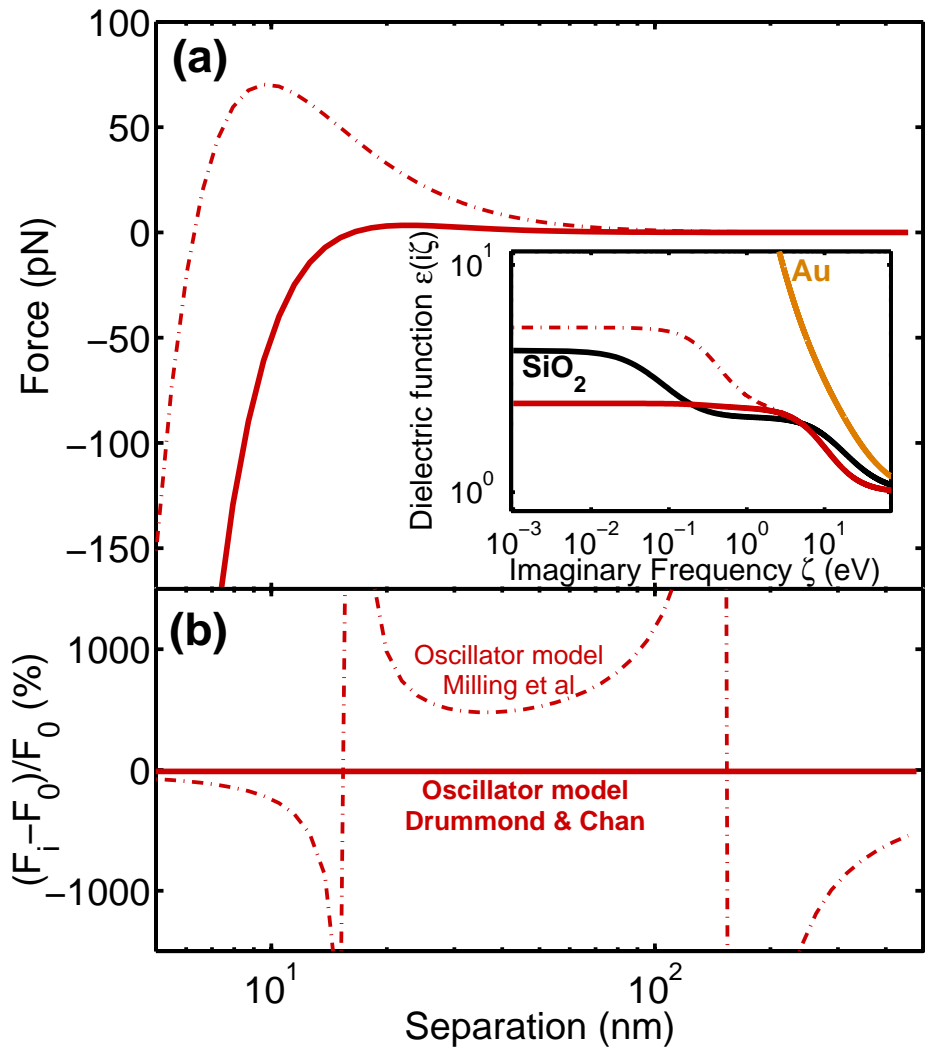

**Figure 9**

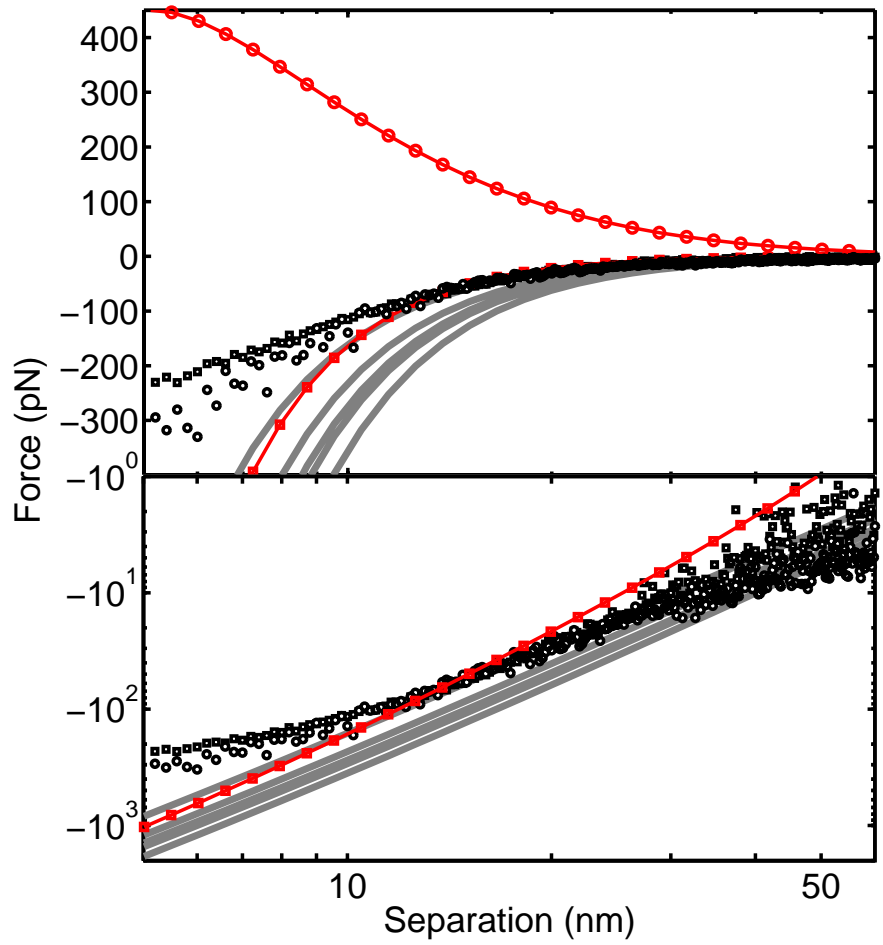

**Figure 10**

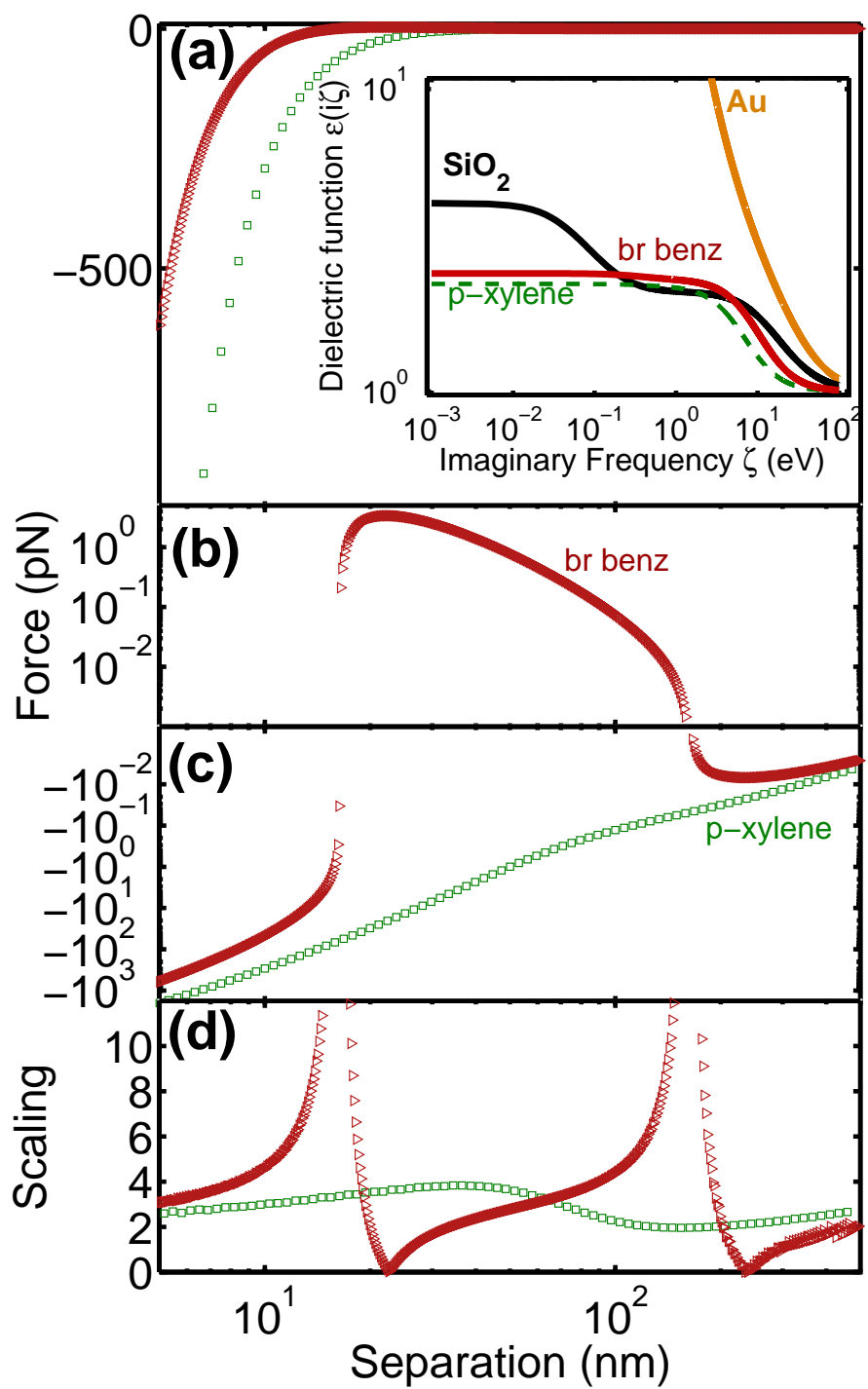

**Figure 11**